\documentstyle[prb,aps]{revtex}

\begin{document}
\draft

\twocolumn[\hsize\textwidth\columnwidth\hsize\csname @twocolumnfalse\endcsname

\title
{
 The 2-D  electron gas at arbitrary spin polarizations
and arbitrary coupling strengths: exchange-correlation energies,
 distribution functions and spin-polarized phases.
}
\author
{
Fran\c{c}ois Perrot, M.W.C. Dharma-wardana\cite{byline1} 
}
\address
{
National Research Council, Ottawa,Canada. K1A 0R6\\
Centre d'Etudes de Bruy\`eres le Ch\^atel, P. O. Box 12,
 91689  Bruy\`eres le Ch\^atel, France
}
\date{24 May 1999}
\maketitle
\begin{abstract}
We use a recent approach
$\lbrack$Phys. Rev. Letters, {\bf 84}, 959 (2000)$\rbrack$
for including Coulomb interactions in quantum systems via
a classical mapping of the pair-distribution functions (PDFs) for
a study of the 2-D electron gas. As in the
3-D case, the ``quantum temperature'' $T_q$ of a {\it classical} 2-D
Coulomb fluid which has the same correlation energy as the
quantum fluid is determined as a function of the density
parameter $r_s$. 
Spin-dependent exchange-correlation
energies are reported.
Comparisons of the spin-dependent pair-distributions
and other calculated properties
 with  any available 2-D
quantum Monte Carlo (QMC) results
show excellent agreement, strongly favouring more recent QMC data.
The  interesting novel
physics brought to light by this study are:
(a) the independently determined quantum-temperatures
for 3-D and 2-D are found to be approximately the {\it same},
(i.e, universal) function of the classical coupling constant $\Gamma$;
(b) the coupling $\Gamma$ increases rapidly with $r_s$
in 2-D, making it comparatively more coupled than in 3-D;
the stronger coupling in 2-D requires bridge corrections to
the hyper-netted-chain method which is adequate in 3-D;
(c) the Helmholtz free energy of spin-polarized and unpolarized
phases have been calculated. The existence of a spin-polarized 2-D liquid 
near $r_s \sim 30$, is found to be a marginal possibility. These results
pertain to clean uniform 2-D electron systems.

\end{abstract}
%\vspace{0.5in}
%\hspace{0.5in} see file /usr/people/chandre/tekstuff/xc2d/ms1.tex 
%
%\hspace{0.5in} date: \today 2-11-2000
\vskip2pc]
\narrowtext

	The past three decades have established the
 two-dimensional electron gas (2DEG) as a goldmine of  new physics and
novel technologies.\cite{ep2ds} The 2DEG, ranging
from a classical to a quantum system, occurs on
liquid-He
surfaces, semiconductor hetero-interfaces, and in
the CuO layers of high-T$_c$  superconductors.
 The  interest in the quantum Hall effect,
 the metal-insulator transition, Kosterlitz-Thouless
 type transitions,\cite{schweigert}
and applications
 to nanotechnology and spintronics
widen its importance.
The exchange-correlation energy ($E_{xc}$) of the 2DEG is 
important in its own right as an input to the
density-functional theory (DFT) of inhomogeneous 2-D systems.

The physics of the 2DEG depends crucially on the  ``coupling parameter''
$\Gamma$ = (potential energy)/(kinetic energy) arising from the Coulomb
interactions.
The $\Gamma$ for the 2DEG at $T=0$ and mean density $n$
is equal to the mean-disk radius $r_s=(\pi n)^{-1/2}$ per electron.
 In the absence of a magnetic
field or disorder effects, the density parameter $r_s$, the
spin polarization $\zeta$ and the temperature
$T$ are the only relevant physical variables.
A major problem in the 2DEG is the need for a theoretical
method applicable at {\it arbitrary } values of $\Gamma$,
$\zeta$ and $T$.
 As in the 3-D electron gas (3DEG), various perturbation
methods, equations-of-motion methods, e.g., 
those of Singwi et al (STLS),\cite{stls}
work for relatively weak coupling, but lead to negative
pair-distribution functions (PDFs)
even at moderate $r_s$.
  In spite of a large
effort, the calculation of the 2DEG $g_{ij}(r)$, where $i,j$ are
spin species,
at arbitrary spin-polarization $\zeta$, $r_s$, and $T$ is
unsolved, except by direct quantum simulation techniques.
Such ``quantum Monte Carlo''(QMC) methods,\cite{ceperley81}
 as well as Feenberg techniques,\cite{feenberg}
 assume a trial wavefunction $\psi=FD$ where $F$ is a correlation
 factor and $D$
is a Slater determinant. The
variational Monte Carlo (VMC) method,\cite{tancep}
and the fixed-node Green-function Monte-Carlo method
 (GFMC) have been applied
to the 2DEG, providing the exchange-correlation energy,
 $E_{xc}(r_s)$ at $T=0$
in a parametrized form.\cite{tancep}
Finite-$T$ systems involve  excited
states as well as the ground state.
While it is easy to get good
$E_{xc}(r_s)$ at $T=0$, $\zeta=0$, the same is not true for most properties
like the PDFs, local-field corrections (LFCs), and xc-energies at finite
$T$, spin polarization $\zeta$ and at strong coupling.

Recently we presented a
computationally simple, conceptually novel {\it classical}
method for calculating the PDFs and other 
properties (e.g, static response) of the 3DEG at
arbitrary coupling, spin polarization and temperature.\cite{prl}
 The method is
based on identifying  
a ``quantum temperature'' $T_q$ such that the correlation
energy of a corresponding {\it classical} Coulomb fluid at $T_q$ is
equal to that of the quantum fluid at $T=0$. Our only ``many-body''
input is a set of values of $E_c(r_s)$ for determining 
$T_q$. Once $T_q(r_s)$ is known, many properties 
inaccessible to standard methods become readily calculable. 
A discussion 
of the $T=0$  and finite-$T$ 3DEG was presented earlier.\cite{prl,prb00} 

A method which works in 3-D does not necessarily work in 
reduced dimensions.
A first objective of this letter is to show that the same ideas
apply to the challenging problems of the 2DEG. In addition, this
study reveals previously unexpected  aspects of the quantum
temperature $T_q$, exposes the long-range nature of the Pauli
 exclusion effect in the 2-D system, and provides new
results for the partially polarized 2-D electron liquid. It
is found to be numerically precise enough, even at $r_s=30$,
 to selectively favour newer QMC calculations.\cite{rapi}
We also examine the stability of spin-polarized
phases of the 2DEG at finite-T and sufficiently large $r_s$.

The PDFs of the 3DEG could be accurately
calculated using the hyper-netted-chain (HNC) approximation.\cite{hncref}
The 2-D system requires that bridge-type cluster
corrections to the  HNC approximation be included.
 A comparison of the
2DEG and 3DEG forms of
 $T_q(r_s)$ reveals that the quantum temperature
is approximately a universal function of the classical-fluid
coupling constant.

The essence of our method is to start from the quantum mechanical
 $g^0(r)$ of the non-interacting problem and build up the interacting
$g(r)$ by classical methods.

Consider a fluid of mean density $n$ containing
two spin species
with concentrations
 $x_i$ = $n_i/n$. 
We deal with the physical temperature $T$ of
the 2DEG, while the temperature $T_{cf}$ of the classical fluid
is $1/\beta$.
Since the leading dependence of the energy on temperature is quadratic,
we assume that $T_{cf}$= ${\surd{(T^2+T_q^2)}}$. This is
clearly valid for $T=0$ and for  high $T$, and was justified
in more detail in ref.~\onlinecite{prb00}.
In this letter the main effort is to
study the 2DEG near $T=0$ and at zero magnetic field  by determining $T_q$.
The equations for the PDFs of a classical
fluid, and the Ornstein-Zernike(OZ) relations are:
\begin{eqnarray}
g_{ij}(r)&=&exp[-\beta \phi_{ij}(r)
+h_{ij}(r)-c_{ij}(r) + B_{ij}(r)]\\
 h_{ij}(r) &=& c_{ij}(r)+
\Sigma_s n_s\int d{\bf r}'h_{i,s}
(|{\bf r}-{\bf r}'|)c_{s,j}({\bf r}')
\label{hnc}
\end{eqnarray}
Here $\phi_{ij}(r)$ is the pair potential between the
species $i,j$. For two electrons this is
just the Coulomb potential $V_{cou}(r)$.
If the spins are parallel, the Pauli
principle prevents  occupation of the same spatial orbital.
As before,\cite{prl}
 we introduce a
``Pauli potential'', ${\cal P}(r)$.
Thus $\phi_{ij}(r)$ becomes ${\cal P}(r)\delta_{ij}+V_{cou}(r)$.
The function $h(r) = g(r)-1$ is related to the
structure factor $S(k)$ by a Fourier transform.
The  $c(r)$ is the ``direct correlation function (DCF)''
of the OZ equations.
The $B_{ij}(r)$  in  Eq. 1 is
the ``bridge'' term due to certain cluster interactions.
If this is neglected,
Eqs. 1-2 form a closed set defining the
HNC approximation.  
The HNC is sufficient for the 3DEG for the
range of $r_s$ studied previously.\cite{prl}
The classical coupling constant
of the 2DEG is found to increase with $r_s$ more rapidly than
 in the 3DEG. Thus,
although ${\cal P}(r)$ restricts clustering effects in the
case of $g_{ii}$, bridge contributions (which contain irreducible
3-body and higher terms) are found to be important for $g_{12}$.
The inclusion of a bridge term into the  HNC equations,
usually via a ``hard-sphere'' model,\cite{rosen} is well
known in 3-D problems, but is less unexplored for 2-D systems.
In this context, it is noteworthy that Kwon et al.
had to include three-body correlations in 
QMC trial functions.\cite{rapi,var}

In the non-interacting system at temperature $T$, 
$n=n_1+n_2$,  $x_i$ = $n_i/n$, the anti-$\|$ $g_{12}^0(r,T)$ is unity while
$$h_{11}^0({\bf r})
=-\frac{1}{n_i^2}\Sigma_{{\bf k}_1,{\bf k}_2}n(k_1)n(k_2)
e^{i({\bf k}_1-{\bf k}_2){\bf \cdot}{\bf r}} \,\,
=\, -[f(r)]^2$$
Here {\bf k}, {\bf r} are 2-D vectors and $n(k)$ is
 the Fermi occupation number at 
the temperature $T$. At $T=0$ $f(r)=2J_1(k_ir)/k_r$ where
$J_1(x)$ is a Bessel function.
 The Pauli exclusion potential is defined by:
\begin{equation}
\beta{\cal{P}}(r)=h_{11}^0(r)-c_{11}^0(r)-ln[g_{11}^0(r)]
\end{equation}
where, e.g., $c^0_{ii}(r)$ is the spin-$\|$ DCF
of the O-Z equation. At $T=0$ it may be shown that 
$\beta{\cal{P}}(r)\sim -2ln(r)$ for $r\to 0$, 
$\beta{\cal{P}}(r)\sim \pi /rk_F$ for $r\to\infty$,
$\beta{\cal{P}}(k)\sim -4\pi /k^2$ for $ k\to\infty$.
Only the
product $\beta{\cal{P}}(r)$ is determined.
 The classical fluid ``temperature'' 
$T_{cf}$=$1/\beta$
is still undefined. The Pauli potential
is a universal function of $rk_F$, and
limited to about a thermal wavelength $\lambda_{th}$
at finite $T$.

	The next step in the method is to use the full 
$\phi_{ij}(r)$,   
and solve the coupled HNC and OZ equations for the binary 
(up and down spins) {\it interacting}$\,$ fluid.
The Coulomb potential $V_{cou}(r)$
for two point-charge electrons is  $1/r$.
However,  an electron at
the temperature $T$ is 
localized to within a thermal wavelength. Thus,  
for the 3DEG we used a ``diffraction
corrected'' form:\cite{minoo},
%\begin{eqnarray}
\begin{equation}
\label{potd}
V_{cou}(r)=(1/r)[1-e^{-rk_{th}}]
\end{equation}
where $k_{th}$ was taken to be  
$ k^0_{th}=(2\pi m^*T_{cf})^{1/2}$ as in Minoo et al.\cite{minoo}
Here $m^*$=1/2 is the reduced mass of the electron {\it pair}.
In the case of the 2DEG we have
determined $k_{th}$ by numerically solving the Schrodinger equation
 for a pair
of 2-D electrons in the potential $1/r$ and calculating the electron density
in each normalized state.\cite{pdwkth}
 Assuming that the diffraction
correction has the same form as before, i.e., $V_d(r)$=exp$(-k_{th}r)/r$,
it was found that $k_{th}/k^0_{th}=1.158T^{0.103}$, where $T$ is in au.

As in the 3DEG, we  need the
 ``quantum temperature'' $T_q$
for each $r_s$ of the 2DEG,  such that the
classical 2D-fluid has the same $E_c(r_s)$ as the quantum fluid.
 Unlike in the
3DEG, an additional complication in the 2DEG is the
need for a bridge function.
In the $\|$-spin case, if the hard-disk radius of the
bridge function is smaller than the
range of the Pauli potential, we may neglect the bridge term.
In the anti-$\|$ case, we need a bridge term
$B_{12}$ explicitly included, say, 
via a hard-disk model.\cite{yr2d}
The same bridge term enters the paramagnetic PDF,
 $g(r)=0.5(g_{11}+g_{12})$ via the $g_{12}$. We can decouple the
 determination of $T_q$ and $B_{12}$
by obtaining $T_q$ from the {\it fully spin polarized}
system as the bridge function is negligible in this case.
Using the analytic
fit of $E_c(r_s,\zeta=1)$ given by 
Tanatar and Ceperley,\cite{tancep}
$t=T_q/T_F$ is found to be fitted by:
\begin{equation}
\label{2dmap}
t=2/(1+0.86413(r_s^{1/6}-1)^2)
\end{equation}
This looks very different from the $T_q(r_s)$ mapping for 
the 3DEG. However, when $T_q$ is plotted as a function of the
{\it classical}  coupling constant $\Gamma_c=1/(T_qr_s)$, (cf. Table I),
 the 3DEG and 2DEG maps are seen to be almost a 
single universal function (Fig. 1.(a)). This result is 
{\it conceptually very interesting}
 since $T_q$ is really a single-parameter representation
of the density-functional correlation energy 
of the electron gas, irrespective of dimensionality.

Unlike in the 3DEG, we need the bridge function for $g_{12}$ in order
to implement the method.
We include only the anti-$\|$ $B_{12}(r)$, and hence
a one-component hard-disk
packing-fraction $\eta$ is needed. The effective coupling constant,
$\Gamma_{eff}=\phi(r_s)/T_{cf}$,
is  a function of $r_sT_q$ and $k_{th}r_s$. That is,
$r_s/t$ and $t^3/r_s$ where $t=T_q/T_F$.
The Gibbs-Bogoliubov  expression for the free energy of the 
one-component hard-disk system has the same dependence.  Also, at large
$r_s$ the packing fraction should tend to a finite limit. Since 
$t$ decreases as $r_s^{-1/3}$,  we consider
the simple functional form 
$\eta=\theta tr_s^{1/3}$.
Here $\theta$ is a ``global'' parameter
 valid for the whole range of $r_s$
and hence the procedure does not involve fitting at each $r_s$. A constant
value, $\theta=0.1175$, proved to be applicable to the range
 $r_s$ =$1-30$  studied
 here. Values of the packing fraction $\eta$,
the QMC and classical-map HNC (CHNC) values of $E_c(r_s,\zeta=0)$
 for the unpolarized 2DEG are
given in Table I. Our $E_c(r_s,\zeta=1)$ agrees with Tanatar-Ceperley 
spin-polarized QMC
by construction. It is common for $\eta$ to approach $\sim$ 0.49 at the
liquid-solid transition. The trend in the value of $\eta$ seen in Table I
is consistent with the finding of Tanatar and Ceperley that the
 2DEG becomes a Wigner solid for $r_s=37\pm5$.

Using the $T_q$ and the hard-disk bridge parameter $\theta$, we
can calculate $E_{c}(r_s,\zeta, T)$ at any $\zeta$. Our
results (at $T=0$) can be presented  via a
 ``polarization
function'':
\begin{eqnarray*}
p(r_s,\zeta) &=& \frac{E_c(r_s,\zeta)-E_c(r_s,0)}{E_c(r_s,1)-E_c(r_s,0)}
\,\, = \frac{\zeta_+^{\alpha(r_s)}-\zeta_-^{\alpha(r_s)}-2}
{2^{\alpha(r_s)}-2}\\
\alpha(r_s)&=&C_1-C_2/r_s+C_3/r_s^{2/3}-C_4/r_s^{1/3}
\;\;\;\;\;\;\;\;\;\;\;\;\;(6)
\end{eqnarray*}
Here $\zeta_{\pm}$=$(1\pm\zeta)$.
In Hartree-Fock, the exponent $\alpha(r_s)$
is a constant, $\alpha=(d+1)/d$ where $d$ is the dimensionality.
Hence, for the 2DEG, the Hartree-Fock value of $\alpha$ is 1.5.
The coefficients $C_1$-$C_4$ are 
1.54039, 0.0305441, 0.296208, and 0.239047 respectively.
A comparison of the 2-D and 3-D $\alpha(r_s)$ is given in Fig.1(b).

Our classical mapping of the 2DEG uses
 the GFMC {\it spin-polarized }
xc-energy fit of Tanatar and Ceperley.\cite{tancep}
Our unpolarized energies (Table I) are in better agreement
with the more recent 
QMC results of Rapisarda et al.\cite{rapi}
In Fig. 2 we compare our PDFs with those of GFMC at $r_s=1$,
5, and 20 given in Tanatar et al, and find satisfactory agreement.

Equation (6) for $p(r_s,\zeta)$ provides a comparison with the
results of Varsano et al.,\cite{var} for the
 correlation energies at partial polarizations,
 for $r_s$=20 and 30. Their
$-E_c(r_s=30,\zeta)$ (au.) are, .01216, .01132, .00987, .00776 at
$\zeta$=.2, .4, .6 and 0.8, while the results from our
fit, eq. (6), are .01219, .01134, .00989, and .00775 respectively.
Varsano et al. suggest that the fully spin-polarized
phase is more stable, `` by an amount close to the statistical
uncertainty''.  Instead of using the fit, we have evaluated the
relevant PDFs accurately at $r_s$=30 and $T$ = 0.
 Varsano et al report a stabilization energy
of -1.185 10$^{-5}$ au., while ours is even smaller and
not significant.

When $r_s$ is large, the kinetic energy is less important;
exchange favours spin-polarization and  correlation
favours the paramagnetic phase.  Figure 3 displays
the Helmholtz free energy  for  $r_s$= 5, 10, 20 and 30.
Stable spin-polarized phases 
in the 3DEG have been reported recently.\cite{ferro3}
Studies related to the metal-insulator transition in the 2DEG, and
numerical studies of 2-D clusters, have increasingly suggested the possibility
of ferromagnetic phases of the 2DEG.\cite{ferro2}
The 2DEG metal-insulator transition is 
influenced by the presence of impurities.
Our results are for the {\it pure, uniform}
system and there seems to be no stable ferromagnetic phase
except near $r_s$=30. 
For 2DEGs in GaAs, $r_s=30$ and $T/E_F$=0.1 corresponds
 to 4.7 10$^{8}$
electrons cm$^{-2}$ and 0.019 Kelvin. The more favourable $m^*$ of
Si brings the densities close to the $\sim 10^{10}$ electrons cm$^{-2}$
range. Details of the finite-$T$ calculations, 
the effect of $\;$impurities on the spin-polarized phase etc.,
will be presented elsewhere.

The computer codes used here may be remotely
accessed by interested workers by visiting our website.\cite{web}

	In conclusion, we present a classical calculation of the
PDFs, correlation energies etc., of the 2DEG for any spin-polarization and
temperature, and find good agreement with
quantum-Monte-Carlo results at $T=0$. The temperature mapping
of the quantum effects obtained for 3-D and 2-D systems is
found to be universal, to within the  accuracy
of the available quantum Monte-Carlo exchange-correlations energies.
New results for
spin-dependent correlation energies and spin polarized
distribution functions are given. A comparison of
the free energies of the  polarized and unpolarized phases
at finite-T is presented.

%\newpage    %%%%%%%%%%%%%%%%%%%%%%%%%%%%%%%%%%%%
%\twocolumn  %%%%%%%%%%%%%%%%%%%%%%%%%%%%%%%%%%%%
%\narrowtext %%%%%%%%%%%%%%%%%%%%%%%%%%%%%%%%%%%%
%

\newpage

\begin{table}
\caption{Relevant parameters and a comparison of the
QMC correlation energies (au.) of Tanatar and Ceperley (TC),
Rapisarda and Senatore (RS), 
with those from the classical map-HNC (CHNC)
for the paramagnetic 2DEG.}

\begin{tabular}{ccccccccc}
$r_s$&$T/T_F$&$\Gamma_c$&$\eta$&$-E_c$(TC)&$-E_c$(RS)&$-E_c$(CHNC)\\
\hline\\
1  & 2.000 & 0.500 &0.235 & 0.10850&- -&0.10838\\
5  & 1.849 & 2.704 &0.372 & 0.04775&0.0489&0.04910\\
10 & 1.682 & 5.946 &0.425 & 0.03034&0.0302&0.03023\\
20 & 1.468 & 13.62 &0.469 & 0.01758&0.0174&0.01739\\
30 & 1.331 & 22.54 &0.486 & 0.01251&0.0124&0.01245\\
\end{tabular}
\label{gamma0}
\end{table}

%
%%%%%%%%%%%%%%%%%%%commenetd out%%%%%%%%%%%%%%%%%%%%%%%%%%
%\begin{figure}
%\caption
%{A comparison of the 2DEG (dashed lines) and 3DEG (solid lines)
% Pauli potentials $\beta {\cal P}(r)$
%at  $T$=0. They are universal functions
%of $rk_F$, where $k_F$ is the Fermi wavevector in 2-D and 3-D.
%}
%\label{fig1}
%\end{figure}
%%%%%%%%%%%%%%%%%%%%%%%%%%%%%%%%%%%%%%%%%%%%%%%%%%%%%%%%%%%%%

\begin{figure}
\caption
{(a) The quantum temperature $T_q$ for the 2DEG (dashes) and the 3DEG
(solid line) as a function
of the classical coupling constant $\Gamma=\beta/r_s$.
(b) The $r_s$ dependence of the 2DEG (dashes) and 3DEG (solid line)
polarization functions
at $T=0$.
}
\label{figGama}
\end{figure}

\begin{figure}
\caption
{ Here the  HNC $g(r)$ are compared with GFMC simulations  of
Tanatar and Ceperley. 
 Solid lines: HNC, boxes: GFMC, 
 Panels (a),(b) unpolarized case,($\zeta=0$), $r_s$=1, and $r_s=20$.
 Panels (c),(d) $r_s$=5 spin-polarized ($\zeta=1$) and unpolarized
 systems.
}
\label{figPDF}
\end{figure}

\begin{figure}
\caption
{ The total Helmholtz free energy of the
unpolarized ($\zeta=0$) and polarized ($\zeta=1$) phases
of the 2DEG near $T=0$ and for $r_s$ = 5, 10, 20, and 30,
as a function of $T/T_F$.
Solid lines: $\zeta=0$; dashes with data points (squares): $\zeta=1$. 
}
\label{figF}
\end{figure}

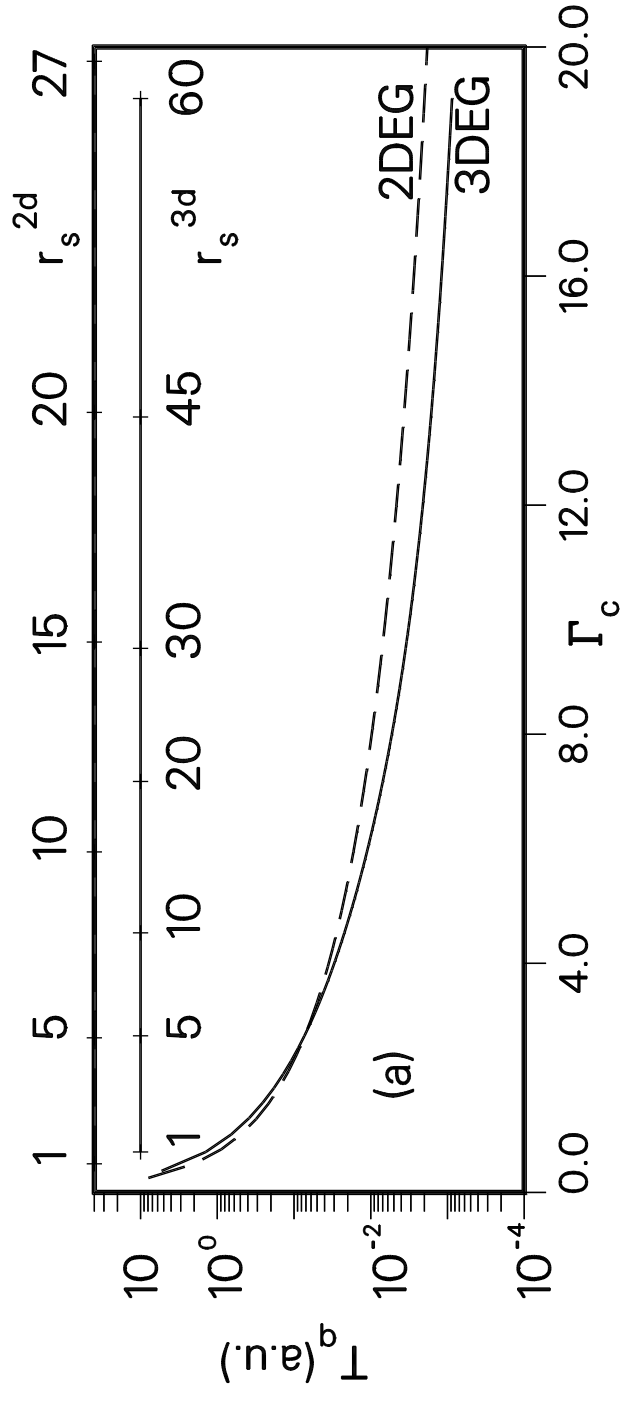
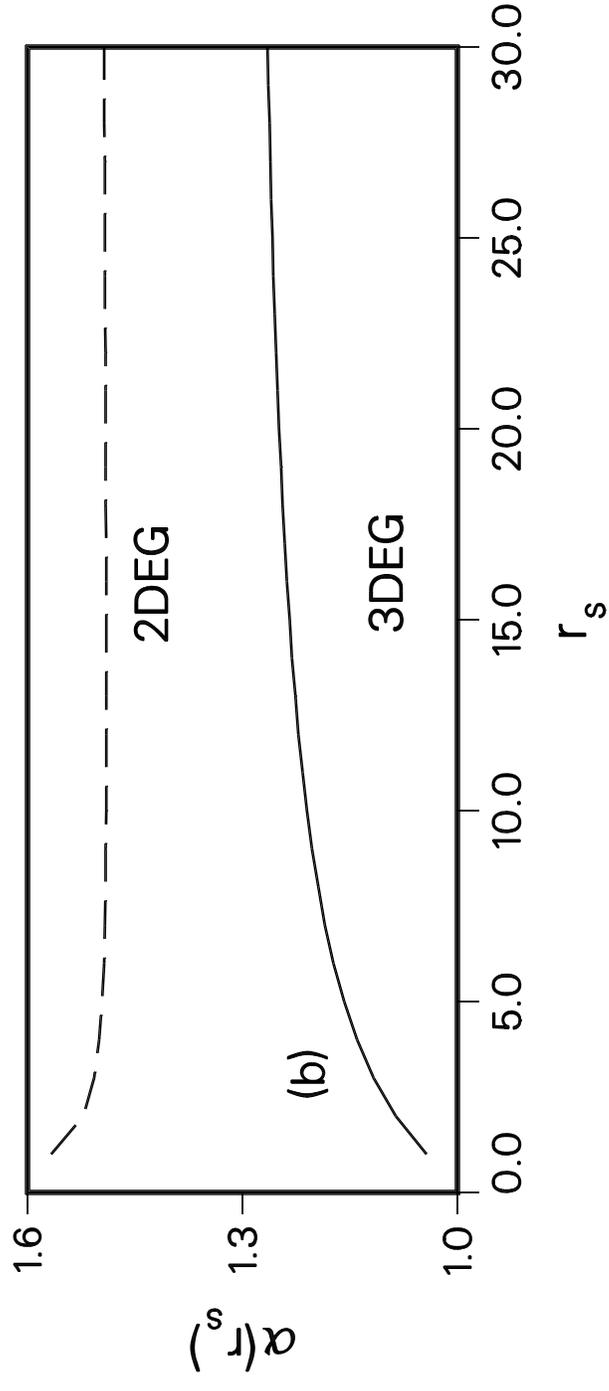
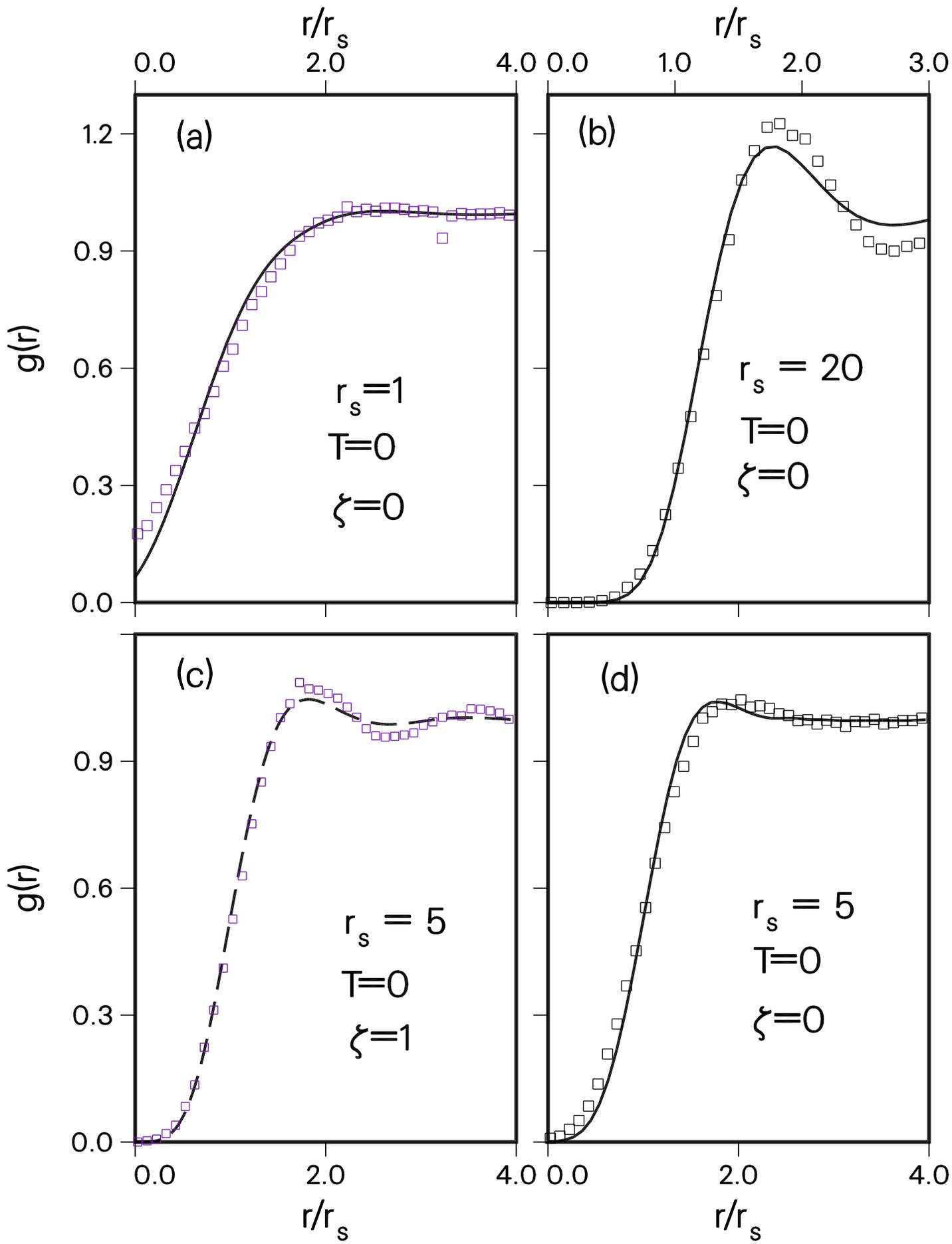
\end{document}